\documentclass[sigconf]{acmart}

\copyrightyear{2026}
\acmYear{2026}
\setcopyright{cc}
\setcctype{by}
\acmConference[ETRA '26]{2026 Symposium on Eye Tracking Research and Applications}{June 01--04, 2026}{Marrakesh, Morocco}
\acmBooktitle{2026 Symposium on Eye Tracking Research and Applications (ETRA '26), June 01--04, 2026, Marrakesh, Morocco}
\acmDOI{10.1145/3797246.3805692}
\acmISBN{979-8-4007-2519-7/2026/06}

\usepackage{booktabs}
\usepackage{multirow}
\usepackage{arydshln}
\usepackage{graphicx}
\usepackage{xcolor}
\usepackage{enumitem}
\usepackage{listings}
\usepackage{subcaption}

\begin{document}

\title{VIVA Stimuli: A Web-Based Platform for Eye Tracking Stimuli}

\author{Suleyman Ozdel}
\authornote{Both authors contributed equally to this work.}
\email{ozdelsuleyman@tum.de}
\affiliation{%
  \institution{Technical University of Munich, Munich Center for Machine Learning}
  \city{Munich}
  \country{Germany}
}

\author{Virmarie Maquiling}
\authornotemark[1]
\email{virmarie.maquiling@tum.de}
\affiliation{%
  \institution{Technical University of Munich, Munich Center for Machine Learning}
  \city{Munich}
  \state{Bavaria}
  \country{Germany}
}

\author{Kadir Burak Buldu}
\email{kadir.buldu@tum.de}
\affiliation{%
  \institution{Technical University of Munich}
  \city{Munich}
  \country{Germany}
}

\author{Yasmeen Abdrabou}
\email{yasmeen.abdrabou@tum.de}
\affiliation{%
  \institution{Technical University of Munich, Munich Center for Machine Learning}
  \city{München}
  \country{Germany}
}

\author{Enkelejda Kasneci}
\email{enkelejda.kasneci@tum.de}
\affiliation{%
  \institution{Technical University of Munich}
  \city{Munich}
  \country{Germany}
}
\affiliation{%
  \institution{Munich Center for Machine Learning}
  \city{Munich}
  \country{Germany}
}

\renewcommand{\shortauthors}{Ozdel and Maquiling et al.}

\begin{abstract}

Reproducibility in eye-tracking research is increasingly important as researchers conduct diverse experiments and seek to validate or replicate findings. However, exact replication remains challenging due to differences in laboratory practices and experimental setups. Inconsistent stimulus presentation can yield divergent metrics from identical oculomotor behavior, yet the stimulus layer remains largely unstandardized. Existing tools often require programming expertise or depend on specific hardware vendors. We introduce VIVA Stimuli, a web-based platform for standardized eye-tracking stimulus presentation. It provides configurable task types, including fixation, smooth pursuit, cognitive load, blink, slippage, content display, and questionnaires within a unified environment. The platform supports any eye-tracking technology, including wearable and screen-based VOG trackers, LFI sensors, and EOG devices. ArUco markers enable synchronization for trackers with scene cameras, while a WebSocket architecture ensures temporal synchronization for those without. A visual experiment flow editor allows protocols to be exported and shared, enabling identical stimulus replication across laboratories.
\end{abstract}



\maketitle

\section{Introduction}
Eye tracking has evolved significantly and is now widely used across many disciplines~\cite{duchowski2017, holmqvist2011,kasneci2024introduction}. It is applied to study reading and information processing in cognitive psychology~\cite{rayner1998}, visual attention in marketing and advertising~\cite{wedel2000}, gaze behavior during driving~\cite{land1994}, usability evaluation in human computer interaction~\cite{poole2006}, learning processes in education~\cite{lai2013,ozdel2025examining}, clinical assessment in neuroscience~\cite{leigh2004}, and interaction in virtual reality~\cite{bozkir2023, maquiling2025imperceptible,ozdel2025exploring}. Across these domains, experiments rely on carefully controlled visual stimuli to evoke fixations and saccades. In practice, however, stimuli are often only partially described or implemented differently across laboratories. Even small design decisions such as target appearance, background luminance, grid spacing, or stimulus duration can influence fixation stability, pupil dynamics, and downstream metrics. As a result, studies that nominally use the same paradigm may produce non comparable findings due to differences in stimulus implementation rather than participants, hardware, or algorithms.

\begin{figure*}[t]
\centering

\begin{subfigure}[t]{0.51\textwidth}
\centering
\includegraphics[width=\textwidth,height=5.5cm,keepaspectratio]{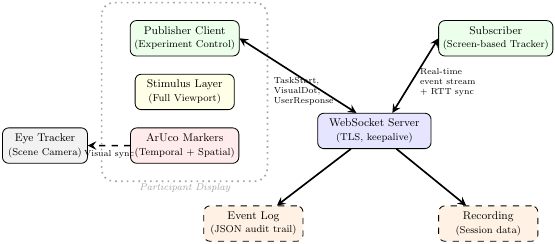}
\caption{System architecture and synchronization.}
\label{fig:architecture}
\end{subfigure}
\hfill
\begin{subfigure}[t]{0.48\textwidth}
\centering
\includegraphics[width=\textwidth,height=5.5cm,keepaspectratio]{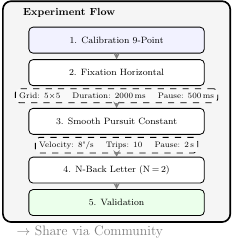}
\caption{Experiment flow editor.}
\label{fig:flow_editor}
\end{subfigure}

\caption{VIVA Stimuli overview. Left: system architecture and synchronization. Right: visual experiment flow editor with shareable protocols.}
\label{fig:overview}

\Description{Left: architecture with publisher, WebSocket server, subscriber, ArUco markers, event log, and eye tracker. Right: experiment flow editor with sequential tasks and configurable parameters.}

\end{figure*}

This reflects a broader reproducibility gap in eye-tracking research. While substantial effort has focused on standardizing analysis methods, event detection algorithms, and reporting guidelines, the stimulus layer remains largely unstandardized. Experimental paradigms are often described only conceptually (e.g., “9-point fixation” or “smooth pursuit”) without fully specifying target appearance, layout or timing. Replication therefore requires re-implementing stimuli from incomplete descriptions, introducing unintended variability. This impacts not only experimental research but also engineering workflows such as benchmarking eye trackers, developing gaze estimation algorithms, and collecting controlled datasets. Existing experiment builders such as PsychoPy~\cite{peirce2019}, OpenSesame~\cite{mathot2012}, and jsPsych~\cite{deleeuw2015} allow flexible stimulus presentation, but tasks must be implemented manually. This results in bespoke scripts whose equivalence across laboratories is difficult to verify. Proprietary vendor tools (e.g., Tobii Pro Lab, SR Research Experiment Builder) offer pre-built tasks but are hardware-locked, closed-source, and tied to specific licenses and devices. Consequently, no hardware-agnostic platform currently enables researchers to define, share, and reproduce eye-tracking stimuli consistently and transparently.

To address this gap, we introduce VIVA Stimuli, a web-based platform for reproducible, shareable, and hardware-agnostic eye-tracking stimuli. The platform offers configurable tasks, including fixation, smooth pursuit, cognitive load, blink, slippage assessment, content display, vergence experiments, calibration routines, and questionnaires. A visual experiment flow editor enables the construction of complete experimental protocols, their sharing within the platform, and their export as machine-readable configurations, allowing identical stimulus sequences and parameters to be reused across laboratories. Synchronization is supported through a WebSocket-based event stream and ArUco marker integration, allowing temporal and spatial alignment with wearable and screen-based eye trackers without vendor-specific APIs. All stimulus events and parameters are stored in a structured event log, enabling the full stimulus sequence to be reconstructed and verified during analysis.


Explicitly defining and logging all stimulus parameters enables VIVA Stimuli to reproduce experiments across different laboratories and hardware setups. This reduces stimulus-related variability when comparing eye trackers, replicating experiments, or collecting datasets for algorithm development, allowing differences in results to be interpreted with fewer stimulus confounds. This work provides a framework for standardized stimulus definition and event logging to support comparable evaluation of eye-tracking metrics and dataset collection across laboratories. The source code and documentation are available at \url{https://gitlab.lrz.de/hctl/viva-stimuli}, and the platform can be accessed at \url{https://hctlsrvc.edu.sot.tum.de/stimuli/}.

\section{Related Work}\label{sec:related}
Several general-purpose experiment frameworks support precise stimulus presentation and flexible task design. PsychoPy~\cite{peirce2019} provides Python-based experiment control with frame-locked timing via OpenGL, while Psychtoolbox~\cite{kleiner2007} delivers millisecond-precise stimulus presentation through MATLAB/Octave with direct hardware access. OpenSesame~\cite{mathot2012} offers a drag-and-drop interface that lowers the programming barrier while maintaining Python extensibility, and jsPsych~\cite{deleeuw2015} enables browser-based experiments via a plugin architecture, popular for online data collection. Gorilla~\cite{anwyl2020} provides a commercial cloud platform combining experimental design with participant recruitment. While flexible and well-validated, these tools require implementing eye tracking tasks from scratch, including fixation targets, pursuit trajectories, synchronization logic, and event logging. As a result, protocols exist as bespoke scripts whose parameter equivalence is difficult to verify without source-code comparison.


\paragraph{Eye Tracking Specific Tools.} Eye tracking manufacturers provide proprietary stimulus software such as Tobii Pro Lab and SR Research Experiment Builder, with pre-built eye tracking tasks. However, these tools are hardware-locked and closed-source: a Tobii Pro Lab protocol cannot run on EyeLink, and neither produces machine-readable files for cross-platform sharing. PyGaze~\cite{dalmaijer2014} addresses hardware lock-in with an open-source interface for multiple trackers via a unified Python API, but still requires custom scripting. Ehinger et al.~[\citeyear{ehinger2019}] developed a standardized test battery evaluating multiple tracker models, demonstrating the value of consistent protocols for benchmarking. However, it was distributed as static image/video sequences rather than a configurable platform, limiting adaptability. Similar static approaches are documented in the EyeLink user manual~\cite{sr2017}.


\begin{figure*}[ht!]
\centering
\includegraphics[width=\textwidth]{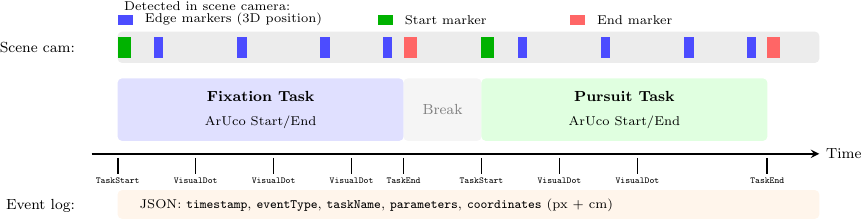}
\Description{Timeline showing task transitions with ArUco markers in the scene camera: blue edge markers visible continuously for 3D position recovery, one green start marker per task, and one red end marker per task, with continuous event logging below.}
\caption{Temporal event sequence during a protocol. Each task begins with one start marker and ends with one end marker, both detected in the scene camera for frame-accurate temporal alignment. Edge markers (blue) provide continuous 3D stimulus position recovery throughout each task. The event log captures every stimulus event with millisecond timestamps and full parameter snapshots.}
\label{fig:timeline}
\end{figure*}

\paragraph{Standardization.} Prior work has focused on improving standardization in eye tracking analysis, reporting, and evaluation. Komogortsev et al.~[\citeyear{komogortsev2010}] proposed standardized metrics and test procedures for fixation and saccade detection. Hessels et al.~[\citeyear{hessels2018}] showed that fixation detection requires careful parameter selection rarely reported in the literature, and introduced the I2MC algorithm as a more robust alternative. Dunn et al.~[\citeyear{dunn2024}] published minimal reporting guidelines covering stimulus properties, apparatus specifications, and procedural details. Hooge et al.~[\citeyear{hooge2022}] evaluated the robustness of wearable trackers under controlled movements, highlighting the need for standardized benchmarks. Niehorster et al.~[\citeyear{niehorster2020}] characterized fixational gaze noise and provided tools for synthesizing realistic noise for algorithm evaluation. While prior work has addressed algorithm standardization and reporting guidelines, fewer efforts have focused on standardizing the stimulus layer for metric benchmarking. Our work complements these efforts by providing infrastructure for controlled and shareable stimulus definition rather than proposing new metrics or detection methods.

\section{Methodology}\label{sec:methodology}

VIVA Stimuli runs in standard browsers without installation. It integrates stimulus tasks, questionnaires, synchronization, and experiment flow management in a single environment. We describe the design tools (Section~\ref{sec:design}), synchronization mechanisms (Section~\ref{sec:sync}), and task library (Section~\ref{sec:tasks}).

\subsection{Experiment Design}\label{sec:design}

Figure~\ref{fig:architecture} illustrates the system architecture, organized around three core principles: the separation of stimulus from experiment control, multi-client synchronization, and comprehensive event logging. The publisher controls the experiment on the participant's display. The WebSocket server provides temporal synchronization for screen-based eye trackers that lack a scene camera. ArUco markers enable synchronization for wearable eye trackers via the scene camera. A JSON event log provides a complete audit trail.

The experiment editor (Figure~\ref{fig:flow_editor}) enables sequencing tasks into protocols without programming. Each step selects a task type from the library and 
allows for independent parameterization. This ensures researchers can apply custom, step-level settings to variables like
dot type, size, color, polarity, timing, and background, so that a single flow can combine different configurations across steps without modifying global settings. Steps can be reordered via drag controls, and breaks can be inserted between tasks. An experiment preset system allows saving, loading, and exporting complete parameter profiles independently of flows, enabling laboratories to share exact settings configurations. A built-in community sharing feature allows researchers to share complete experiments directly within the VIVA Stimuli environment, 
enabling users to import and run the exact same protocol configuration with all step-specific customizations perfectly preserved. 

\subsection{Synchronization and 3D Stimulus Positioning}\label{sec:sync}

Eye-tracking research uses diverse technologies, including VOG-based wearable trackers (e.g., Pupil Labs, Tobii Pro Glasses), screen-based trackers (e.g., Tobii Pro, EyeLink), LFI sensors, and EOG devices, which differ in whether they provide a scene camera. The platform supports two complementary synchronization mechanisms and spatial positioning for 3D stimulus recovery, enabling integration across tracker types without manufacturer-specific APIs. The framework focuses on providing structured temporal markers and event logs to support downstream alignment and benchmarking. 

\subsubsection{Subscriber Synchronization.} 
Screen-based eye trackers (e.g., Tobii Pro, EyeLink), LFI sensors, and EOG devices lack a scene camera and cannot detect visual markers. For these systems, the platform provides WebSocket-based synchronization. A single publisher controls the session, while subscribers receive a real-time event stream for temporal alignment. Structured JSON messages include millisecond timestamps, round-trip time (RTT)–based latency estimates, and task-level records of all stimulus-related events and associated parameters. The configuration is serialized at each task boundary, recording active parameters per trial. Appendix~\ref{App_a} shows representative fixation events.

\subsubsection{ArUco Temporal Synchronization.} Each task is assigned a dedicated pair of ArUco marker IDs~\cite{garrido2014}, one start marker displayed when the task begins and one end marker displayed when it concludes. These markers are briefly displayed at task transitions, creating detectable events in the scene camera that enable frame-aligned temporal synchronization.
The precision of this alignment is inherently bounded by the scene camera's frame rate and the reliability of the marker detection algorithm.
Each task type is allocated a dedicated ID range, providing unique identification that allows automated post-processing of the data. Figure~\ref{fig:timeline} illustrates the temporal event sequence during a task transition.

\subsubsection{3D Stimulus Positioning via Edge Markers.} 
\begin{figure*}[t!]
\centering

\begin{subfigure}[c]{0.55\textwidth}
\centering
\includegraphics[width=\textwidth]{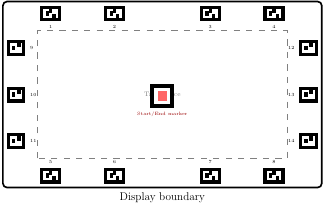}
\caption{Edge marker layout. Fourteen ArUco markers define the screen perimeter and the internal task space.}
\label{fig:edge_markers}
\end{subfigure}\hfill
\begin{subfigure}[c]{0.43\textwidth}
\centering
\includegraphics[width=\textwidth]{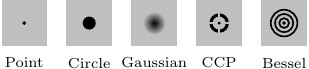}
\caption{The five fixation target types (dark-on-gray polarity)~\cite{niehorster2025best}.}
\label{fig:fixation_targets}
\end{subfigure}

\caption{Experimental setup components.}
\label{fig:combined_setup}
\Description{Left: edge marker layout showing 14 ArUco markers around the screen perimeter. Right: five fixation target types (Point, Circle, Gaussian, CCP, Bessel) in dark-on-gray polarity.}
\end{figure*}

Beyond temporal synchronization, 14 edge markers are permanently positioned around the screen perimeter at known viewport coordinates during tasks (Figure~\ref{fig:edge_markers}). Four markers span the top edge, four the bottom, and three line each edge. They enable an external camera to recover the 3D display plane via PnP (Perspective-n-Point) pose estimation, providing pixel-to-centimeter conversion without manual measurement. For eye trackers with a scene camera, this allows fully recreating the 3D position of each stimulus, which is essential for gaze estimation algorithms and evaluating existing methods.


\begin{table*}[ht!]
\caption{Stimulus task categories and configurable parameters.}
\label{tab:tasks}
\footnotesize
\setlength{\tabcolsep}{4pt}
\begin{tabular}{@{}p{2cm}p{4.5cm}p{6.5cm}@{}}
\toprule
\textbf{Category} & \textbf{Task Types} & \textbf{Key Configurable Parameters} \\
\midrule

\textbf{Fixation} 
& Horizontal, Vertical, Random (Saccade) 
& Grid size, duration, pause, target type, stimulus mode, polarity \\

\textbf{Smooth Pursuit} 
& Constant, Accelerating, Random, Circular, Wandering 
& Velocity, direction, trips, acceleration, duration, viewing distance \\

\textbf{Cognitive Load} 
& N-Back (Letter, Shape, Sound); Stroop (Simple, Complex, Fast); Arrow tracking (3 levels) 
& Difficulty level, trials, timing, grid size \\

\textbf{Blink} 
& Long, Short Interval 
& Interval duration, cue count \\

\textbf{Slippage} 
& 12-step head movement protocol 
& Step duration, grid size \\

\textbf{Content} 
& Text, Image, Video 
& Display duration, custom media \\

\textbf{Vergence} 
& Beam splitter depth stimulus 
& Depth positions, marker mode \\

\textbf{Calibration} 
& Calibration (1/3/5/9-pt), Validation 
& Point count, target type, duration \\

\textbf{Questionnaires} 
& Demographics, NASA-TLX, BFI-10 
& -- \\

\bottomrule
\end{tabular}
\end{table*}

\subsection{Task Library}\label{sec:tasks}
The task library includes six stimulus categories: fixation, smooth pursuit, cognitive load, blink, slippage, and content display (text, image, video). It also provides calibration and validation routines, a vergence task for beam splitter setups, and validated questionnaires within a unified eye-tracking environment. The task library is designed to support common benchmarking and metric-evaluation scenarios, including fixation-precision estimation, smooth-pursuit accuracy analysis, slippage-robustness testing, cognitive-load pupillometry, and content-based gaze-behavior studies. Table~\ref{tab:tasks} summarizes the task library. Each task exposes configurable parameters that researchers can adapt while maintaining configuration-level reproducibility through shared protocol exports.

\subsubsection{Fixation (Random Saccade Task).} 
Tasks present stationary targets on configurable grids (up to 20$\times$20) in horizontal, vertical, or pseudorandom modes. Positions are reported in pixel and physical coordinates and computed relative to edge ArUco markers, allowing adaptation to different screens while preserving physical spacing. Timing parameters are configurable. Each mode supports three stimulus types: a standard dot, an E stimulus (a rotated letter “E” used to verify fixation at the stimulus location), and a shrinking dot that progressively reduces angular subtense. The pseudorandom mode elicits saccades~\cite{leigh2004}. Dedicated ArUco marker ranges uniquely label grid rows for synchronization. The choice of fixation target significantly influences data quality~\cite{thaler2013, niehorster2025best}. We implement five perceptually distinct targets (Figure~\ref{fig:fixation_targets}).
The \textbf{Point} is a minimal filled dot that provides the smallest possible fixation anchor. 
The \textbf{Circle} is a filled disc and represents the most common fixation target used in eye-tracking studies. 
The \textbf{Gaussian} target consists of a radial gradient from the target color to the background, with a configurable sigma, that avoids hard luminance boundaries. 
The \textbf{CCP (Crosshair--Circle--Point)} target is a bullseye pattern composed of a filled annulus, a crosshair gap, and a center dot, following~\cite{thaler2013}, which showed that bullseye targets can reduce fixation dispersion compared to simple dots. 
Finally, the \textbf{Bessel} target uses concentric rings with configurable spatial frequencies, providing multi-scale spatial structure to support stable fixation.
All five targets support dark-on-gray and light-on-dark polarity modes, custom RGB colors, and opacity control. The polarity mode affects the entire background, ensuring consistent adaptation conditions. These parameters influence pupillometry and fixation stability~\cite{holmqvist2011, thaler2013}, yet are rarely standardized across studies.

\subsubsection{Smooth Pursuit.}
Tasks move a target along controlled trajectories in five modes. \textbf{Constant} presents linear motion at fixed velocity.  \textbf{Accelerating} increases velocity across successive trips.  \textbf{Random} samples velocity and acceleration from configurable ranges to generate unpredictable dynamics. \textbf{Circular} follows an elliptical orbit with adjustable radius, duration, and direction, supporting vestibular and neurological assessment. Lastly, \textbf{Wandering} produces a smooth random walk with boundary avoidance for naturalistic motion. All modes use frame-synchronized animation with configurable inter-trip pauses. Target velocity is scaled to viewing distance and monitor size to maintain consistent angular velocity across participants and setups.

\subsubsection{Cognitive Load.} 
Tasks assess pupil dilation and fixation under varying mental effort~\cite{chen2014, beatty1982, krejtz2018}. \textbf{N-Back}~\cite{kirchner1958} is implemented in letter, shape, and auditory modalities with configurable difficulty and timing; responses and accuracy are logged with millisecond precision. \textbf{Stroop}~\cite{stroop1935} includes congruent, incongruent and fast incongruent variants with configurable trials and timing, supporting voice responses when available. \textbf{Arrow tracking} presents an arrow moving through an $8\times8$ grid, where participants follow the arrow and count the number of direction changes. Three difficulty levels control the number of turns (3, 7, and 11 for easy, medium, and hard), increasing working memory and attentional demands~\cite{wu2024pupil}.

\subsubsection{Blink.} Tasks present auditory cues at controlled intervals in two variants for baseline blink rate characterization: a long-interval variant with alternating ``open'' and ``close'' text and beep cues, and a short-interval variant with beep-only cues. This allows triggering instant blinks and enables correct labeling of completely closed eyelid states.

\subsubsection{Slippage Assessment.} Implements a 12-step standardized protocol for quantifying head-mounted tracker slippage, inspired by \citeauthor{niehorster2020impact}~[\citeyear{niehorster2020impact, niehorster2018expect}]. The sequence progresses through: (1)~a fixation grid, (2)~central fixation, (3)~central fixation during speech production, (4)~eyebrow raise and furrow, (5 to 7)~head rotations in roll, pitch, and yaw with photographic instructions, (8 to 10)~micro-translations in lateral, vertical, and depth axes, (11)~return to original position, and (12)~a post-perturbation fixation grid to quantify residual offset. Each step has configurable duration and a unique ArUco marker for temporal segmentation. This enables direct evaluation of the eye trackers against slippage.

\subsubsection{Calibration and Validation.} The platform provides 1-, 3-, 5-, and 9-point calibration routines, each with an integrated 4-point validation phase, all with ArUco-marked transitions. Calibration points use the selected fixation target type for consistency with experimental stimuli. These routines support external calibration pipelines rather than replacing manufacturer-specific calibration.

\subsubsection{Content Display.} Tasks present text passages, images, or video clips with configurable durations and dedicated ArUco markers for temporal segmentation. Text passages support reading studies where eye movements indicate cognitive processing~\cite{rayner1998, just1976}. Pre-loaded content includes texts, images, and videos, with custom media upload available.

\subsubsection{Vergence.} The vergence task enables depth-axis experiments analogous to VR vergence studies. 
The task requires a beam splitter that aligns a phone and a screen along the z-axis. Either display can present ArUco markers for scene camera depth detection, or event logs with manual measurements provide equivalent depth information.

\subsubsection{Questionnaires.} An embedded questionnaire module supports radio, checkbox, text, number, and Likert-scale fields.   Three instruments are included: a 21-item demographics survey covering participant characteristics and eye-tracking experience, the NASA Task Load Index (NASA-TLX)~\cite{hart1988} in both weighted and raw variants with six workload dimensions, and the BFI-10 personality inventory~\cite{rammstedt2007} with automatic dimension scoring. Additionally, custom questionnaires can be created with multiple-choice, Likert-scale, and text input questions. Questionnaires can be integrated into the experiment flow as steps alongside stimulus tasks.  All responses, along with user IDs and session metadata, are timestamped and saved in a single unified event log, maintaining a complete audit trail.

\section{Discussion}\label{sec:discussion}

For researchers developing and evaluating eye-tracking metrics, stimulus variability is a critical but often overlooked confound. Differences in fixation target design, pursuit trajectories, or timing parameters can significantly influence measured accuracy, precision, and robustness metrics. By formalizing stimulus definitions and explicitly logging them, VIVA Stimuli provides infrastructure that enables more interpretable metric comparisons across systems and datasets. In practice, this reduces setup time and eliminates the need to re-implement tasks, enabling researchers to deploy standardized protocols quickly without additional development effort.

VIVA Stimuli addresses this gap by making stimulus definitions explicit, UI-configurable, portable, and machine-readable. Instead of textual descriptions, complete stimulus configurations are encoded, logged, and shareable. This shifts stimulus design from an implicit implementation detail to an auditable experimental component. As a result, the same protocol configurations can be executed across laboratories without re-implementation, reducing stimulus variability as a confound and improving comparability of results. Beyond replication, this enables standardized benchmarking and dataset collection. When stimulus parameters are fixed and reproducible, differences in measured performance can be attributed with greater confidence to the eye tracker, the participant population, or the analysis pipeline. This is particularly relevant for evaluating gaze estimation algorithms, comparing hardware systems, and constructing shared datasets.

\subsubsection{Hardware Agnostic Synchronization.} 

An important strength of the platform is hardware independence. By combining ArUco-based scene-camera alignment with WebSocket-based temporal synchronization, the same stimulus protocol can operate across wearable and screen-based trackers, LFI sensors, and EOG devices. For systems with a scene camera, start and end markers enable frame-level alignment between stimulus and gaze logs. For systems without one, real-time event streaming provides direct temporal synchronization. In addition, edge markers support spatial calibration by recovering the screen’s 3D pose, enabling precise stimulus localization for gaze estimation and benchmarking. Together, these mechanisms decouple stimulus presentation from manufacturer-specific APIs while preserving temporal and spatial consistency across eye-tracking systems.

\subsubsection{Comparison with Existing Tools.} 
General-purpose builders (PsychoPy~\cite{peirce2019}, Psychtoolbox~\cite{kleiner2007}, jsPsych~\cite{deleeuw2015}, OpenSesame~\cite{mathot2012}) require scripting for each task and lack shareable protocol formats. Hardware-agnostic middleware such as PyGaze~\cite{dalmaijer2014} simplifies eye tracker integration but still requires programming. Proprietary tools (Tobii Pro Lab, SR Research Experiment Builder) offer pre-built tasks but are hardware-locked and closed-source. 
Considering these limitations, VIVA Stimuli simplifies experiment implementation and unifies the data collection process, removing the need to manage multiple data sources or handle unsynchronized values. The framework addresses practical challenges during experiment setup and execution while supporting complete experiment designs. It provides a consistent structure for conducting experiments across different eye-tracking systems with or without a scene camera. This standardization enables reliable data collection for eye-tracking metric evaluation and cross-laboratory benchmarking.

\subsubsection{Limitations and Future Work.}

As a browser-based system, stimulus timing is constrained by the web rendering pipeline. At 60\,Hz, \texttt{requestAnimationFrame} provides frame-level granularity (16.7\,ms), and JavaScript scheduling may introduce additional jitter. For many fixation and cognitive paradigms, this level of precision is sufficient relative to task durations and measurement noise~\cite{holmqvist2011}, and smooth pursuit tasks remain stable through frame-synchronized velocity updates. However, paradigms requiring sub-frame accuracy should be validated on target hardware or implemented using dedicated frameworks such as PsychoPy, which is not web-based but runs as a desktop application and achieves frame-locked precision via direct OpenGL rendering~\cite{peirce2019}. Synchronization accuracy depends on the properties of the eye tracker and scene camera, and ArUco-based alignment is bounded by camera frame rate and detection reliability.

The platform focuses on stimulus presentation and does not process gaze data, preserving hardware independence but preventing gaze-contingent paradigms. Photometric calibration is not included; luminance-sensitive experiments such as pupillometry~\cite{mathot2018} require independent display calibration. Formal timing validation remains future work. Despite these constraints, sharing explicit stimulus protocols can reduce stimulus-related variability in cross-laboratory benchmarking and dataset collection.

\section{Conclusion}\label{sec:conclusion}

We presented VIVA Stimuli, a web-based framework for standardized stimulus protocols and logging for 
eye-tracking metric evaluation and cross-lab benchmarking. The platform complements analysis standardization~\cite{komogortsev2010, andersson2017} and reporting guidelines~\cite{dunn2024} by making the stimulus layer explicit and portable. Future work will improve community sharing and extend task coverage.

\begin{acks}
This project is supported by the Chips Joint Undertaking (Chips JU) and its members, including top-up funding by Denmark, Germany, Netherlands, Sweden, under grant agreement No.\ 101139942.
\end{acks}


\bibliographystyle{ACM-Reference-Format}
\bibliography{references}
\appendix
\section{APPENDIX: Json Event Log}
\label{App_a}
\begin{figure}[ht]
\centering
\begin{lstlisting}[
  basicstyle=\ttfamily\tiny,
  frame=single,
  breaklines=true,
  numbers=none,
  xleftmargin=2pt,
  xrightmargin=2pt,
  aboveskip=4pt,
  belowskip=4pt
]
{"timestamp":1740500100000,
 "eventType":"TaskStart",
 "taskName":"HorizontalFixation",
 "details":{"markerId":49,
   "currentSettings":{"fixationDotType":"CCP",
     "fixationDotSize":2.5,
     "fixationDotPolarity":"dark-on-gray",
     "gridRows":5,"gridCols":5,
     "displayDuration":1500,
     "pauseDuration":300,
     "screenWidthCm":60.5,
     "screenHeightCm":33.5}}}
\end{lstlisting}
\vspace{4pt}
\begin{lstlisting}[
  basicstyle=\ttfamily\tiny,
  frame=single,
  breaklines=true,
  numbers=none,
  xleftmargin=2pt,
  xrightmargin=2pt,
  aboveskip=4pt,
  belowskip=4pt
]
{"timestamp":1740500101500,
 "eventType":"VisualDot",
 "details":{"xPx":384,"yPx":216,
   "xCm":12.1,"yCm":6.7,
   "dotType":"CCP",
   "dotPolarity":"dark-on-gray",
   "taskName":"HorizontalFixation"}}
\end{lstlisting}
\caption{Representative event log entries for a horizontal fixation task. Each entry includes a timestamp, event type, and payload with both pixel and physical coordinates.}
\label{lst:eventlog}
\Description{JSON event log showing TaskStart and VisualDot entries stacked vertically with timestamps, parameters, and coordinates.}
\end{figure}
\end{document}